arXiv:1609.04234v1 [stat.ME] 14 Sep 2016

# Two New Tests for Equality of Several Covariance Functions

Jia Guo and Jin-Ting Zhang

*National University of Singapore*

**Abstract**

In this paper, we propose two new tests for testing the equality of the covariance functions of several functional populations, namely a quasi GPF test and a quasi $F_{\max}$ test. The asymptotic random expressions of the two tests under the null hypothesis are derived. We show that the asymptotic null distribution of the quasi GPF test is a chi-squared-type mixture whose distribution can be well approximated by a simple scaled chi-squared distribution. We also adopt a random permutation method for approximating the null distributions of the quasi GPF and $F_{\max}$ tests. The random permutation method is applicable for both large and finite sample sizes. The asymptotic distributions of the two tests under a local alternative are investigated and they are shown to be root-n consistent. Simulation studies are presented to demonstrate the finite-sample performance of the new tests against three existing tests. They show that our new tests are more powerful than the three existing tests when the covariance functions at different time points have different scales. An illustrative example is also presented.

## 1 Introduction

In recent decades, increasing attention has been paid to functional data whose observations are functions, such as curves, surfaces, or images. Such a kind of data arises frequently in various research and industrial areas. How to analyze these functional data becomes a hot topic and novel methodologies to deal with them are in great demand. Many classical statistical methods for multivariate data, such as principal component analysis and canonical correlation analysis among others, have been extended to satisfy this need. Among these methods, hypothesis testing for functional data also attracts increasing interests from researchers. Most popular hypothesis testing problems are inferences concerning means or covariances.

It is well known that in the classical analysis of variance (ANOVA), the $F$-test is a widely used tool

Jia Guo (E-mail: jia.guo@u.nus.edu) is PhD candidate, Jin-Ting Zhang (E-mail: stazjt@nus.edu.sg) is Associate Professor, Department of Statistics and Applied Probability, National University of Singapore, Singapore 117546. The work was financially supported by the National University of Singapore Academic Research grant R-155-000-164-112.

which uses the ratio of the sum of squares between subjects (SSB) and the sum of squares due to errors (SSE) as its test statistic. That is $F = \frac{\text{SSB}/(k-1)}{\text{SSE}/(n-k)}$ where $n$ and $k$ are the sample size and the number of groups respectively, SSB and SSE measure the variations explained by the factors involved in the analysis and the variations due to measurement errors. Due to its robustness, the $F$-test is often recommended in practice. In the functional data analysis, we can define SSB and SSE for each time point and denote them as SSB($t$) and SSE($t$) respectively. The test statistic of the pointwise $F$-test described by Ramsay and Silverman (2005) can be defined as $F(t) = \frac{\text{SSB}(t)/(k-1)}{\text{SSE}(t)/(n-k)}$ which is a natural extension of the classical $F$-test to the field of functional data analysis; see more details in Section 2 below. However, this test is time-consuming and cannot give a global conclusion. To overcome this difficulty, Cuevas et al. (2004) proposed an ANOVA test based on the $L^2$-norm of SSB($t$), i.e., the numerator of the pointwise $F$-test statistic but its asymptotic null distribution of the test statistic is not given. Zhang (2013) further investigated this test statistic which is called the $L^2$-norm based test and showed that its null distribution is asymptotically a $\chi^2$-type mixture. Instead of only using the numerator of the pointwise $F$-test, Zhang and Liang (2013) studied a GPF test which is obtained via globalizing the pointwise $F$-test with integration. Alternatively, the pointwise $F$-test can be globalized via using its maximum value as a test statistic, resulting in the so-called $F_{\max}$-test as described by Cheng et al. (2012). It is shown that the $F_{\max}$ test is powerful when the functional data are highly correlated and the GPF test is powerful when the functional data are less correlated. Besides its importance in functional ANOVA problems, the pointwise $F$-test can also be applied in functional linear models. In fact, Shen and Faraway (2004) considered an $F$-type test to compare two nested linear models and studied its null distribution. Their test relies on the integrated residual sum of squares proposed in Faraway (1997). Based on their work, Zhang (2011) studied the asymptotic power of this $F$-type test and extended it to a general linear hypothesis testing (GLHT) problem.

In the above, we can see that the pointwise $F$-test is quite useful and powerful in functional data analysis and it can be globalized to yield the so-called $GPF$ and $F_{\max}$ tests among others. This paper aims to develop a similar pointwise test for the equality of the covariance functions of several functional populations, namely, the equal-covariance function (ECF) testing problem. This task is quite challenging and novel since the pointwise $F$-test is usually defined only for the one-way ANOVA problem or the regression analysis as mentioned above. In fact, it is very difficult to define such a pointwise $F$-test for the ECF testing problem. Instead, we can only mimic the basic idea of the pointwise $F$-test and define a pointwise quasi $F$-test for the ECF testing problem as we shall do in Section 2 below. Based on this pointwise quasi $F$-test, we construct two new globalized tests, namely, a quasi GPF test and a quasi $F_{\max}$ test. The asymptotic random expressions of the test statistics under both the null and alternative hypotheses are derived. To approximate the null distribution of the quasi GPF test, two methods are proposed. One applies the Welch-Satterthwaite $\chi^2$-approximation and the other applies the random permutation method. For the quasi $F_{\max}$ test, we only use the random permutation method. Like the classical $F$-test, these two new tests are scale-invariant. In addition, we show, via simulation studies, that our new tests are more powerful than three existing tests when the covariance functions at different time points have different scales.

The paper is organized as follows. The main results are presented in Section 2. The simulation studies are presented in Section 3. A real data example is given in Section 4. The technical proofs of our main results are presented in the Appendix.

## 2 Main Results

Let $y_{i1}(t), y_{i2}(t), \cdots, y_{in_i}(t)$, $i = 1, 2, \cdots, k$ be $k$ independent functional samples over a given finite time interval $\mathcal{T} = [a, b]$, $-\infty < a < b < \infty$, which satisfy

$$\begin{array}{c} y_{ij}(t) = \eta_i(t) + v_{ij}(t),\ j = 1, 2, \cdots, n_i, \\ v_{i1}(t), v_{i2}(t), \cdots, v_{in_i}(t) \stackrel{i.i.d.}{\sim} \text{SP}(0, \gamma_i);\ i = 1, 2, \cdots, k, \end{array} \tag{2.1}$$

where $\eta_1(t), \eta_2(t), \cdots, \eta_k(t)$ model the unknown group mean functions of the $k$ samples, $v_{ij}(t),\ j = 1, 2, \cdots, n_i,\ i = 1, 2, \cdots, k$ denote the subject-effect functions, and $\gamma_i(s, t),\ i = 1, 2, \cdots, k$ are the associated covariance functions. Throughout this paper, we assume that $\text{tr}(\gamma_i) < \infty$ and $\eta_i(t) \in \mathcal{L}^2(\mathcal{T})$, $i = 1, 2, \cdots, k$, where $\mathcal{L}^2(\mathcal{T})$ denotes the Hilbert space formed by all the squared integrable functions over $\mathcal{T}$ with the inner-product defined as $< f, g >= \int_{\mathcal{T}} f(t)g(t)dt,\ f,\ g \in \mathcal{L}^2(\mathcal{T})$. It is often of interest to test the equality of the $k$ covariance functions:

$$H_0 : \gamma_1(s, t) \equiv \gamma_2(s, t) \equiv \cdots \equiv \gamma_k(s, t), \quad \text{for all } s, t \in \mathcal{T}. \tag{2.2}$$

For convenience, we refer to the above problem as the $k$-sample equal-covariance function (ECF) testing problem for functional data.

Based on the given $k$ functional samples (2.1), the group mean functions $\eta_i(t),\ i = 1, 2, \cdots, k$ and the covariance functions $\gamma_i(s, t),\ i = 1, 2, \cdots, k$ can be unbiasedly estimated as

$$\begin{array}{rcl} \hat{\eta}_i(t) & = & \bar{y}_i(t) = n_i^{-1} \sum_{j=1}^{n_i} y_{ij}(t),\ i = 1, 2, \cdots, k, \\ \hat{\gamma}_i(s, t) & = & (n_i - 1)^{-1} \sum_{j=1}^{n_i} [y_{ij}(s) - \bar{y}_i(s)][y_{ij}(t) - \bar{y}_i(t)],\ i = 1, 2, \cdots, k. \end{array} \tag{2.3}$$

It is easy to show that $\hat{\gamma}_i(s, t),\ i = 1, 2, \cdots, k$ are independent and $\text{E}\hat{\gamma}_i(s, t) = \gamma_i(s, t),\ i = 1, 2, \cdots, k$. Further, the estimated subject-effect functions can be written as

$$\hat{v}_{ij}(t) = y_{ij}(t) - \bar{y}_i(t),\ j = 1, 2, \cdots, n_i;\ i = 1, 2, \cdots, k. \tag{2.4}$$

When the null hypothesis (2.2) holds, let $\gamma(s, t)$ denote the common covariance function of the $k$ samples. It can be estimated by the following pooled sample covariance function

$$\hat{\gamma}(s, t) = \sum_{i=1}^{k} (n_i - 1)\hat{\gamma}_i(s, t)/(n - k), \tag{2.5}$$

where $\hat{\gamma}_i(s,t),\ i=1,2,\cdots,k$ are given in (2.3).

The tests we shall propose are inspired by the GPF test of Zhang and Liang (2013) and the $F_{\max}$-test of Cheng et al. (2012). Both of them are based on the pointwise $F$-test as mentioned in the introduction. To better understand how we shall define our new tests, we first review the GPF and $F_{\max}$-tests. These two tests are designed to test the one-way ANOVA for functional data, i.e., to test if the $k$ mean functions are equal: $H_0: \eta_1(t)=\eta_2(t)=\cdots=\eta_k(t)$. For this end, Zhang and Liang (2013) first defined the pointwise sum of squares between groups (SSB) and the pointwise sum of squares due to errors (SSE):

$$\mathrm{SSB}(t)=\sum_{i=1}^k n_i[\hat{\eta}_i(t)-\hat{\eta}(t)]^2,\ \ \mathrm{SSE}(t)=\sum_{i=1}^k\sum_{j=1}^{n_i}[y_{ij}(t)-\hat{\eta}_i(t)]^2, \tag{2.6}$$

where $\hat{\eta}(t)=\sum_{i=1}^k n_i\hat{\eta}_i(t)/n$ denotes the pooled sample mean function of the $k$ functional samples. Then the pointwise $F$-test statistic can be defined as

$$F_n(t)=\frac{\mathrm{SSB}(t)/(k-1)}{\mathrm{SSE}(t)/(n-k)}, \tag{2.7}$$

where and throughout $n=\sum_{i=1}^k n_i$ denotes the total sample size. The test statistics of the GPF and $F_{\max}$ tests are then given respectively by

$$T_n=\int_{\mathcal{T}} F_n(t)dt,\ \ F_{\max}=\sup_{t\in\mathcal{T}} F_n(t). \tag{2.8}$$

Our new test statistics can be defined similarly but they are based on a pointwise quasi $F$-test. For the ECF testing problem (2.2), we first define the pointwise sum of squares between groups (SSB) and sum of squares due to errors (SSE):

$$\mathrm{SSB}(s,t)=\sum_{i=1}^k (n_i-1)[\hat{\gamma}_i(s,t)-\hat{\gamma}(s,t)]^2,\ \ \mathrm{SSE}(s,t)=\sum_{i=1}^k\sum_{j=1}^{n_i}[\hat{v}_{ij}(s)\hat{v}_{ij}(t)-\hat{\gamma}_i(s,t)]^2,\ \ s,t\in\mathcal{T},$$

where $\hat{\gamma}(s,t)$, the pooled sample covariance function of the $k$ functional samples as defined in (2.5), $\hat{\gamma}_i(s,t)$, the $i$-th sample covariance function, and $\hat{v}_{ij}(s)\hat{v}_{ij}(t)$ play the roles of $\hat{\mu}(t)$, $\hat{\mu}_i(t)$ and $y_{ij}(t)$ in (2.6) respectively. Then the pointwise quasi $F$-test statistic for testing (2.2) can be defined as

$$F_n(s,t)=\frac{\mathrm{SSB}(s,t)/(k-1)}{\mathrm{SSE}(s,t)/(n-k)},\ \ s,t\in\mathcal{T}, \tag{2.9}$$

which may not have an $F$-distribution and hence $F_n(s,t)$ should not be called a pointwise $F$-test statistic. Then the test statistic obtained via integrating the pointwise quasi $F$-test statistic may be called a quasi GPF test statistic and the test statistic obtained via taking the supremum of the pointwise quasi $F$-test statistic may be called a quasi $F_{\max}$ test statistic. That is, the test statistics of the quasi GPF and $F_{\max}$ tests are then given respectively by

$$T_n=\int_{\mathcal{T}}\int_{\mathcal{T}} F_n(s,t)dsdt,\ \ F_{\max}=\sup_{s,t\in\mathcal{T}} F_n(s,t). \tag{2.10}$$

Notice that when the null hypothesis is valid, it is expected that both $T_n$ and $F_{\max}$ will be small and otherwise large.

For further study, let $\varpi_i\left[(s_1,t_1),(s_2,t_2)\right]$ denote the covariance function between $v_{i1}(s_1)v_{i1}(t_1)$ and $v_{i1}(s_2)v_{i1}(t_2)$. Then we have

$$\varpi_i\left[(s_1,t_1),(s_2,t_2)\right] = \mathrm{E}\{v_{i1}(s_1)v_{i1}(t_1)v_{i1}(s_2)v_{i1}(t_2)\} - \gamma_i(s_1,t_1)\gamma_i(s_2,t_2). \tag{2.11}$$

When $\gamma_i(s,t)$ does not depend on $i$, i.e., when $H_0$ holds, we use $\gamma(s,t)$ to denote the common covariance function, and define

$$\varpi\left[(s_1,t_1),(s_2,t_2)\right] = n^{-1}\sum_{i=1}^{k} n_i \mathrm{E}\{v_{i1}(s_1)v_{i1}(t_1)v_{i1}(s_2)v_{i1}(t_2)\} - \gamma(s_1,t_1)\gamma(s_2,t_2). \tag{2.12}$$

The natural estimator for $\varpi\left[(s_1,t_1),(s_2,t_2)\right]$ is

$$\hat{\varpi}\left[(s_1,t_1),(s_2,t_2)\right] = n^{-1}\sum_{i=1}^{k}\sum_{j=1}^{n_i} \hat{v}_{ij}(s_1)\hat{v}_{ij}(t_1)\hat{v}_{ij}(s_2)\hat{v}_{ij}(t_2) - \hat{\gamma}(s_1,t_1)\hat{\gamma}(s_2,t_2). \tag{2.13}$$

When the samples are Gaussian, a consistent estimator of $\varpi\left[(s_1,t_1),(s_2,t_2)\right]$ is given by

$$\hat{\varpi}[(s_1,t_1),(s_2,t_2)] = \hat{\gamma}(s_1,s_2)\hat{\gamma}(t_1,t_2) + \hat{\gamma}(s_1,t_2)\hat{\gamma}(s_2,t_1). \tag{2.14}$$

To derive the asymptotic random expressions of $T_n$ and $F_{\max}$, we impose the following assumptions:

Assumption A

1. The $k$ samples are Gaussian.
2. As $n \to \infty$, the $k$ sample sizes satisfy $n_i/n \to \tau_i \in (0,1),\ i = 1,2,\cdots,k$.
3. The variance functions are uniformly bounded. That is, $\rho_i = \sup_{t\in\mathcal{T}} \gamma_i(t,t) < \infty,\ i = 1,2,\cdots,k$.

Assumption A2 requires that the $k$ sample sizes tend to $\infty$ proportionally.

Before we state the main results, we give an alternative expression of $\mathrm{SSB}(s,t)$ which is helpful for deriving the main results about the quasi GPF and $F_{\max}$ tests. For any $s,t \in \mathcal{T}$, $\mathrm{SSB}(s,t)$ can be expressed as

$$\mathrm{SSB}(s,t) = \mathbf{z}_n(s,t)^T[\mathbf{I}_k - \mathbf{b}_n\mathbf{b}_n^T/(n-k)]\mathbf{z}_n(s,t), \tag{2.15}$$

where

$$\mathbf{z}_n(s,t) = [z_1(s,t), z_2(s,t), \cdots, z_k(s,t)]^T,$$

with

$$\begin{aligned} z_i[s,t] &= \sqrt{n_i-1}[\hat{\gamma}_i(s,t) - \gamma(s,t)],\ i = 1,2,\cdots,k, \\ \mathbf{b}_n &= [\sqrt{n_1-1}, \sqrt{n_2-1}, \cdots, \sqrt{n_k-1}]^T. \end{aligned}$$

Since $\mathbf{b}_n^T\mathbf{b}_n/(n-k)=1$, it is easy to verify that $\mathbf{I}_k-\mathbf{b}_n\mathbf{b}_n^T/(n-k)$ is an idempotent matrix with rank $k-1$. In addition, as $n\to\infty$, we have

$$\mathbf{I}_k-\mathbf{b}_n\mathbf{b}_n^T/(n-k)\to\mathbf{I}_k-\mathbf{b}\mathbf{b}^T,\text{with }\mathbf{b}=[\sqrt{\tau_1},\sqrt{\tau_2},\cdots,\sqrt{\tau_k}]^T, \tag{2.16}$$

where $\tau_i, i=1,2,\cdots,k$ are given in Assumption A2. Note that $\mathbf{I}_k-\mathbf{b}\mathbf{b}^T$ in (2.16) is also an idempotent matrix of rank $k-1$, which has the following singular value decomposition:

$$\mathbf{I}_k-\mathbf{b}\mathbf{b}^T=\mathbf{U}\begin{pmatrix}\mathbf{I}_{k-1} & \mathbf{0}\\ \mathbf{0}^T & 0\end{pmatrix}\mathbf{U}^T, \tag{2.17}$$

where the columns of $\mathbf{U}$ are the eigenvectors of $\mathbf{I}_k-\mathbf{b}\mathbf{b}^T$. We now have the following theorem.

**Theorem 1.** *Under Assumptions A1∼A3 and the null hypothesis (2.2), as* $n\to\infty$, *we have* $T_n\xrightarrow{d}T_0$ *with*

$$\begin{aligned}T_0 &\stackrel{d}{=} \int_{\mathcal{T}}\int_{\mathcal{T}}(k-1)^{-1}\sum_{i=1}^{k-1}\omega_i^2(s,t)dsdt\\ &\stackrel{d}{=} (k-1)^{-1}\sum_{r=1}^{\infty}\lambda_rA_r,\ \ A_r\stackrel{i.i.d.}{\sim}\chi^2_{k-1},\end{aligned} \tag{2.18}$$

*and* $F_{\max}\xrightarrow{d}F_0$ *with*

$$F_0\stackrel{d}{=}sup_{s,t\in\mathcal{T}}\{(k-1)^{-1}\sum_{i=1}^{k-1}\omega_i^2(s,t)\}, \tag{2.19}$$

*where* $\omega_1(s,t),\omega_2(s,t),\cdots,\omega_{k-1}(s,t)\stackrel{i.i.d.}{\sim}GP(0,\gamma_\omega)$ *with*

$$\gamma_\omega[(s_1,t_1),(s_2,t_2)]=\varpi[(s_1,t_1),(s_2,t_2)]/\sqrt{\varpi[(s_1,t_1),(s_1,t_1)]\varpi[(s_2,t_2),(s_2,t_2)]}, \tag{2.20}$$

*and* $\varpi[(s_1,t_1),(s_2,t_2)]$ *is defined in (2.12), and* $\lambda_r,\ r=1,2,\cdots,\infty$ *are the decreasing-ordered eigenvalues of* $\gamma_\omega[(s_1,t_1),(s_2,t_2)]$.

By Theorem 1, $\omega_i(s,t),\ i=1,2,\cdots,k\stackrel{i.i.d.}{\sim}GP(0,\gamma_\omega)$ which are known except $\gamma_\omega[(s_1,t_1),(s_2,t_2)]$. The covariance function $\gamma_\omega[(s_1,t_1),(s_2,t_2)]$ can be estimated by

$$\hat{\gamma}_\omega[(s_1,t_1),(s_2,t_2)]=\frac{\hat{\varpi}[(s_1,t_1),(s_2,t_2)]}{\sqrt{\hat{\varpi}[(s_1,t_1),(s_1,t_1)]\hat{\varpi}[(s_2,t_2),(s_2,t_2)]}} \tag{2.21}$$

where $\hat{\varpi}[(s_1,t_1),(s_2,t_2)]$ is given in (2.13) or (2.14).

Theorem 1 says that the asymptotic distribution of $T_n$ is the same as that of a $\chi^2$-type mixture. Therefore we can approximate its distribution using the well-known Welch-Satterthwaite $\chi^2$-approximation. That is, we approximate the null distribution of $T_n$ using that of a random variable

$$R\stackrel{d}{=}\beta\chi^2_d \tag{2.22}$$

via matching the first two moments of $T_n$ and $R$. By some simple algebra, we have

$$\beta=\frac{\text{tr}(\gamma_\omega^{\otimes 2})}{(k-1)\text{tr}(\gamma_\omega)},\ d=\frac{(k-1)\text{tr}^2(\gamma_\omega)}{\text{tr}(\gamma_\omega^{\otimes 2})}, \tag{2.23}$$

where

$$\begin{aligned}\mathrm{tr}(\gamma_\omega) &= \int_{\mathcal{T}}\int_{\mathcal{T}}\gamma_\omega\left[(s,t),(s,t)\right]dsdt = (b-a)^2,\\ \mathrm{tr}(\gamma_\omega^{\otimes 2}) &= \int_{\mathcal{T}}\int_{\mathcal{T}}\int_{\mathcal{T}}\int_{\mathcal{T}}\gamma_\omega^2\left[(s_1,t_1),(s_2,t_2)\right]ds_1dt_1ds_2dt_2.\end{aligned}$$

The quasi GPF test can be implemented provided that the parameters $\beta$ and $d$ are properly estimated. For the given $k$ samples, we can obtain the following naive estimators of $\beta$ and $d$ via replacing $\gamma_\omega\left[(s_1,t_1),(s_2,t_2)\right]$ with its estimator $\hat{\gamma}_\omega\left[(s_1,t_1),(s_2,t_2)\right]$ as given in (2.21) in the expressions (2.23):

$$\hat{\beta} = \frac{\mathrm{tr}(\hat{\gamma}_\omega^{\otimes 2})}{(k-1)(b-a)^2},\quad \hat{d} = \frac{(k-1)(b-a)^4}{\mathrm{tr}(\hat{\gamma}_\omega^{\otimes 2})}, \tag{2.24}$$

where $\hat{\gamma}_\omega\left[(s_1,t_1),(s_2,t_2)\right]$ is given in (2.21). Then we have

$$T_n \sim \hat{\beta}\chi^2_{\hat{d}} \text{ approximately}, \tag{2.25}$$

so that the quasi GPF test can be conducted accordingly.

**Theorem 2.** *Under Assumptions A1∼A3 and the null hypothesis (2.2), as $n\to\infty$, we have $\hat{\beta}\xrightarrow{p}\beta$, $\hat{d}\xrightarrow{p} d$ and $\hat{C}_\alpha \xrightarrow{p} \tilde{C}_\alpha$ where $\hat{C}_\alpha = \hat{\beta}\chi^2_{\hat{d}}(\alpha)$ is the estimated critical value of $T_n$ and $\tilde{C}_\alpha = \beta\chi^2_d(\alpha)$ is the approximate theoretical critical value of $T_n$.*

Theorem 2 shows that the naive estimators $\hat{\beta}$ and $\hat{d}$ converge in probability to their underlying values and thus the estimated $100\alpha$-quantile converges to the theoretical $100\alpha$-quantile. The naive estimators are simple to implement and easy to compute. However, it requires that the group sample sizes are large so that the asymptotic results of Theorem 1 are valid.

Alternatively, we can adopt the following random permutation method for approximating the null distribution of the quasi GPF and $F_{\max}$ tests. This random permutation method is applicable for both large and small sample sizes. Let

$$v_{ij}^*(t),\ j=1,2,\cdots,n_i;\ i=1,2,\cdots,k, \tag{2.26}$$

be the $k$ permuted samples generated from the estimated subject-effect functions given in (2.4). That is, we first permute the estimated subject-effect functions $\hat{v}_{ij}(t), j=1,2,\cdots,n_i; i=1,2,\cdots,k$ and then use the first $n_1$ functions as $v_{1j}^*(t), j=1,2,\cdots,n_1$ and use the next $n_2$ functions as $v_{2j}^*(t), j=1,2,\cdots,n_2$ and so on. It is obvious that given the original $k$ functional samples (2.1), the $k$ permuted samples (2.26) are i.i.d with mean function 0 and covariance function $\frac{n-k}{n}\hat{\gamma}(s,t)$, where $\hat{\gamma}(s,t)$ is the pooled sample covariance function given in (2.5). Then the permuted test statistics of the quasi GPF and $F_{\max}$ tests based on the $k$ permuted samples can be obtained similarly as we defined $T_n$ and $F_{\max}$ based on the $k$ original functional samples (2.1). That is, the permuted test statistics can be obtained as

$$T_n^* = \int_{\mathcal{T}}\int_{\mathcal{T}} F_n^*(s,t)dsdt,\quad F_{\max}^* = sup_{s,t\in\mathcal{T}}F_n^*(s,t)$$

where

$$\begin{aligned} F_n^*(s,t) &= \tfrac{SSB^*(s,t)/(k-1)}{SSE^*(s,t)/(n-k)}, \\ \mathrm{SSB}^*(s,t) &= \textstyle\sum_{i=1}^k (n_i-1)[\hat{\gamma}_i^*(s,t)-\hat{\gamma}^*(s,t)]^2, \\ \mathrm{SSE}^*(s,t) &= \textstyle\sum_{i=1}^k \sum_{j=1}^{n_i} [\hat{v}_{ij}^*(s)\hat{v}_{ij}(t)-\hat{\gamma}_i^*(s,t)]^2, \end{aligned}$$

with

$$\hat{\gamma}_i^*(s,t) = (n_i-1)^{-1}\textstyle\sum_{j=1}^{n_i} \hat{v}_{ij}^*(s)\hat{v}_{ij}^*(t), i=1,2,\cdots,k,$$
$$\hat{\gamma}^*(s,t) = \textstyle\sum_{i=1}^k (n_i-1)\hat{\gamma}_i^*(s,t)/(n-k).$$

The permuted upper $100\alpha$-percentiles $C_{1\alpha}^*$ and $C_{2\alpha}^*$ of $T_n^*$ and $F_{\max}^*$ can then be obtained via repeating the above random permutation process a large number of times.

Let $C_{1\alpha}$ and $C_{2\alpha}$ denote the upper $100\alpha$-percentiles of $T_0$ and $F_0$ respectively, where $T_0$ and $F_0$ are the limit random variables of $T_n$ and $F_{\max}$ under the null hypothesis $H_0$ as defined in Theorem 1. The following theorem shows that the permutation test statistics admit the same limit random expressions of the original test statistics and hence the associated critical values $C_{1\alpha}^*$ and $C_{2\alpha}^*$ will tend to $C_{1\alpha}$ and $C_{2\alpha}$ in distribution as $n \to \infty$. Thus we can use the critical values $C_{1\alpha}^*$ and $C_{2\alpha}^*$ to conduct the quasi GPF and $F_{\max}$ tests.

**Theorem 3.** *Under Assumptions A1∼A3 and the null hypothesis (2.2), as $n \to \infty$, we have $T_n^* \xrightarrow{d} T_0$, $F_{\max}^* \xrightarrow{d} F_0$ and $C_{1\alpha}^* \xrightarrow{d} C_{1\alpha}$, $C_{2\alpha}^* \xrightarrow{d} C_{2\alpha}$.*

We now study the asymptotic powers of the quasi GPF and $F_{\max}$ tests under the following local alternative:

$$H_1 : \gamma_i(s,t) = \gamma(s,t) + (n_i-1)^{-1/2} d_i(s,t),\ i = 1,2,\cdots,k, \tag{2.27}$$

where $d_1(s,t), d_2(s,t), \cdots, d_k(s,t)$ are some fixed bivariate functions, independent of $n$ and $\gamma(s,t)$ is some covariance function. This local alternative will tend to the null hypothesis in a root-$n$ rate and hence it is difficult to detect. First of all, we derive the alternative distribution of the quasi $F_{\max}$ test in Theorem 4 and that of the quasi GPF test in Theorem 5 below.

**Theorem 4.** *Under Assumptions A1∼A3 and the local alternative (2.27), as $n \to \infty$, we have $F_{\max} \xrightarrow{d} F_1$ with*

$$F_1 \stackrel{d}{=} sup_{s,t\in\mathcal{T}}\Big\{(k-1)^{-1}\sum_{i=1}^{k-1}[\omega_i(s,t)+\zeta_{\varpi i}(s,t)]^2\Big\},$$

*where $\omega_1(s,t), \omega_2(s,t), \cdots, \omega_{k-1}(s,t) \stackrel{i.i.d.}{\sim} GP(0,\gamma_\omega)$ as in Theorem 1 and $\zeta_{\varpi i}(s,t),\ i = 1,2,\cdots,k-1$ are the $(k-1)$ components of $\boldsymbol{\zeta}_\varpi(s,t) = (\mathbf{I}_{k-1},\mathbf{0})\mathbf{U}^T\mathbf{d}(s,t)/\sqrt{\varpi[(s,t),(s,t)]}$ with $\mathbf{U}$ given in (2.17), $\varpi[(s,t),(s,t)]$ given in (2.12) and $\mathbf{d}(s,t) = [d_1(s,t), d_2(s,t),\cdots,d_k(s,t)]^T$ with its entries given in (2.27).*

Let $\lambda_r$, $r = 1,2,\cdots,\infty$ be the eigenvalues of $\gamma_\omega[(s_1,t_1),(s_2,t_2)]$ with only the first $m$ eigenvalues being positive and $\phi_r(s,t)$, $r = 1,2,\cdots,\infty$ are the associated eigenfunctions.

**Theorem 5.** *Under Assumptions A1∼A3 and the local alternative (2.27), as $n \to \infty$, we have $T_n \xrightarrow{d} R_1$ with*

$$\begin{aligned} R_1 &\stackrel{d}{=} (k-1)^{-1}\int_{\mathcal{T}}\int_{\mathcal{T}} ||\mathbf{x}(s,t)||^2 dsdt = (k-1)^{-1}\sum_{i=1}^{k-1}\int_{\mathcal{T}}\int_{\mathcal{T}} x_i^2(s,t)dsdt \\ &\stackrel{d}{=} (k-1)^{-1}[\sum_{r=1}^{m}\lambda_r A_r + \sum_{r=m+1}^{\infty}\delta_r^2], \end{aligned}$$

*where $A_r \sim \chi^2_{k-1}(\lambda_r^{-1}\delta_r^2)$, $r = 1, 2, \cdots, m$, are independent, $\mathbf{x}(s,t) = [x_1(s,t), x_2(s,t), \cdots, x_{k-1}(s,t)]^T \sim GP_{k-1}(\boldsymbol{\zeta}_\varpi(s,t),\ \gamma_\omega \mathbf{I}_{k-1})$ with $\boldsymbol{\zeta}_\varpi(s,t)$ defined in Theorem 4, and $\delta_r^2 = ||\int_{\mathcal{T}}\int_{\mathcal{T}} \boldsymbol{\zeta}_\varpi(s,t)\phi_r(s,t)dsdt||^2$, $r = 1, 2, \cdots, \infty$.*

Theorem 6 states the asymptotic normality of the quasi GPF test under the local alternative (2.27). Theorems 7 and 8 show that the quasi GPF and $F_{\max}$ tests are root-$n$ consistent. In these three theorems, the quantities $\delta_r^2, r = 1, 2, \cdots$ are defined in Theorem 5.

**Theorem 6.** *Under Assumptions A1∼A3 and the local alternative (2.27), as $\max_r \delta_r^2 \to \infty$, we have*

$$\frac{T_n - E(T_n)}{\sqrt{Var(T_n)}} \xrightarrow{d} N(0,1).$$

**Theorem 7.** *Under Assumptions A1∼A3 and the local alternative (2.27), as $\max_r \delta_r^2 \to \infty$, the quasi GPF test has asymptotic power 1. That is, $P(T_n > C_\alpha) \to 1$ where $C_\alpha$ can be $\hat{C}_\alpha = \hat{\beta}\chi^2_{\hat{d}}(\alpha)$, the estimated critical value of $T_n$, or $C^*_{1\alpha}$, the estimated upper 100$\alpha$-percentile of $T_n$ using the random permutation method.*

**Theorem 8.** *Under Assumptions A1∼A3 and the local alternative (2.27), as $n \to \infty$, the power of the quasi $F_{\max}$ test $P(F_{\max} \geq C^*_{2\alpha})$ will tend to 1 as $\max_r \delta_r^2 \to \infty$ where $C^*_{2\alpha}$ is the estimated upper 100$\alpha$-percentile of the random permuted test statistic $F^*_{\max}$.*

In the proof of Theorem 8, we shall use the following relationship between the quasi $F_{\max}$ test statistic and the quasi GPF test statistic defined in (2.10) :

$$T_n = \int_{\mathcal{T}}\int_{\mathcal{T}} F_n(s,t)dsdt \leq (b-a)^2 F_{\max}, \tag{2.28}$$

where we use the fact that $\mathcal{T} = [a, b]$. It then follows that

$$P(F_{\max} \geq C^*_{2\alpha}) \geq P(T_n \geq (b-a)^2 C^*_{2\alpha}). \tag{2.29}$$

However, we cannot compare the values of $(b-a)^2 C^*_{2\alpha}$ and the upper 100$\alpha$-percentile of the quasi GPF test statistic $T_n$. Thus, the expression (2.29) does not guarantee that the quasi $F_{\max}$ test is more powerful than the quasi GPF test. To compare the powers of these two tests, some simulation studies are then needed.

# 3 Simulation Studies

For the ECF testing problem, Guo et al. (2016) studied an $L^2$-norm based test. They proposed to approximate the null distribution of the $L^2$-norm based test statistic using a naive method, a bias-reduced method, and a random permutation method. The associated tests can be represented by $L^2_{nv}, L^2_{br}$ and $L^2_{rp}$ respectively. When the functional data are Gaussian, $L^2_{br}$ and $L^2_{rp}$ are comparable and they outperform $L^2_{nv}$ in general. For the ECF testing problem, Zhou et al. (2016) proposed a so-called $T_{\max,rp}$-test using the supremum value of the sum of the squared differences between the group sample covariance functions and the associated pooled sample covariance function. When functional data are highly correlated, they showed that the $T_{\max,rp}$-test has higher powers than $L^2_{nv}, L^2_{br}$ and $L^2_{rp}$. Since we can approximate the null distribution of the quasi GPF test using a naive method and a random permutation method, the associated quasi GPF tests are denoted as $\text{GPF}_{nv}$ and $\text{GPF}_{rp}$ respectively. Similarly, we denote the quasi $F_{\max}$ test with the random permutation method to approximate the associated null distribution by $F_{\max,rp}$ for simplicity. In this section, we present some simulation studies, aiming to compare $\text{GPF}_{nv}, \text{GPF}_{rp}$ and $F_{\max,rp}$ against $L^2_{br}, L^2_{rp}$ and $T_{\max,rp}$. We exclude $L^2_{nv}$ since its performance is not as good as $L^2_{br}, L^2_{rp}$ and $T_{\max,rp}$. In this section, we shall present three different simulation studies for three different goals.

## 3.1 Data Generating

We use the following model to generate $k$ functional samples:

$$\begin{gathered} y_{ij}(t) = \eta_i(t) + v_{ij}(t),\ \eta_i(t) = \mathbf{c}_i^T[1, t, t^2, t^3]^T,\ v_{ij}(t) = \mathbf{b}_{ij}^T \mathbf{\Psi}_i(t),\ t \in [0,1], \\ \mathbf{b}_{ij} = [b_{ij1}, b_{ij2}, \cdots, b_{ijq}]^T,\ b_{ijr} \stackrel{d}{=} \sqrt{\lambda_r} z_{ijr},\ r = 1, 2, \cdots, q; \end{gathered} \tag{3.1}$$

$j = 1, 2, \cdots, n_i,\ i = 1, 2, \cdots, k,$ where $\eta_i(t),\ i = 1, 2, \cdots, k$ are the group mean functions with the parameter vectors $\mathbf{c}_i = [c_{i1}, c_{i2}, c_{i3}, c_{i4}]^T,\ i = 1, 2, \cdots, k,\ \mathbf{\Psi}_i(t) = [\psi_{i1}(t), \psi_{i2}(t), \cdots, \psi_{iq}(t)]^T$ is a vector of $q$ basis functions $\psi_{ir}(t),\ t \in [0, 1],\ r = 1, 2, \cdots, q,$ the variance components $\lambda_r,\ r = 1, 2, \cdots, q$ are positive and decreasing in $r$, and the number of the basis functions $q$ is an odd positive integer and the random variables $z_{ijr},\ r = 1, 2, \cdots, q;\ j = 1, 2, \cdots, n_i;\ i = 1, 2, \cdots, k$ are i.i.d. with mean 0 and variance 1. Then we have the group mean functions $\eta_i(t) = c_{i1} + c_{i2}t + c_{i3}t^2 + c_{i4}t^3$, $i = 1, 2, \cdots, k$ and the group covariance functions

$$\gamma_i(s, t) = \mathbf{\Psi}_i(s)^T diag(\lambda_1, \lambda_2, \cdots, \lambda_q) \mathbf{\Psi}_i(t) = \sum_{r=1}^{q} \lambda_r \psi_{ir}(s) \psi_{ir}(t), i = 1, 2, \cdots, k.$$

In the simulations, the design time points for all the functions $y_{ij}(t),\ j = 1, 2, \cdots, n_i,\ i = 1, 2, \cdots, k$ are assumed to be the same and are specified as $t_j = (j-1)/(J-1),\ j = 1, 2, \cdots, J,$ where $J$ is some positive integer.

We next specify the model parameters in (3.1). We choose the group number $k = 3$. To specify the group mean functions $\eta_1(t), \eta_2(t), \cdots, \eta_k(t)$, we set $\mathbf{c}_1 = [1, 2.3, 3.4, 1.5]^T$ and $\mathbf{c}_i = \mathbf{c}_1 + (i-1)\delta\mathbf{u},\ i = 2, 3,$ where the tuning parameter $\delta$ specifies the differences $\eta_i(t) - \eta_1(t),\ i = 2, 3$, and the constant vector $\mathbf{u}$ specifies the direction of these differences. We set $\delta = 0.1$ and $\mathbf{u} = [1, 2, 3, 4]^T/\sqrt{30}$ which is a unit vector. Then we specify the covariance functions $\gamma_i(s, t), i = 1, 2, \cdots, k$. For simplicity, we set $\lambda_r = a\rho^{r-1},\ r = 1, 2, \cdots, q$, for some $a > 0$ and $0 < \rho < 1$. Notice that the tuning parameter $\rho$ not only determines the decay rate of $\lambda_1, \lambda_2, \cdots, \lambda_q$, but also determines how the simulated functional data are correlated: when $\rho$ is close to 0, $\lambda_1, \lambda_2, \cdots, \lambda_q$ will decay very fast, indicating that the simulated functional data are highly correlated; and when $\rho$ is close to 1, $\lambda_r,\ r = 1, 2, \cdots, q$ will decay very slowly, indicating that the simulated functional data are nearly uncorrelated. The functions $\psi_{ir}(t),\ i = 1, 2, 3;\ r = 1, 2, \cdots, q$ in the above model (3.1) are carefully specified. First of all, let $\phi_1(t) = 1,\ \phi_{2r}(t) = \sqrt{2}sin(2\pi rt),\ \phi_{2r+1}(t) = \sqrt{2}cos(2\pi rt),\ t \in [0, 1],\ r = 1, 2, \cdots, (q - 1)/2$ to be a vector of $q$ orthonormal basis functions $\boldsymbol{\phi}(t) = [\phi_1(t), \phi_2(t), \cdots, \phi_q(t)]^T$, and specify $\psi_{ir}(t) = \phi_r(t),\ r = 1, 3, 4, \cdots, q$ and $\psi_{i2}(t) = \phi_2(t) + (i - 1)\omega$ respectively where $\omega$ is some constant. It can be seen the covariance functions are

$$\gamma_i(s, t) = \gamma_1(s, t) + (i - 1)\lambda_2[\phi_2(s) + \phi_2(t)]\omega + (i - 1)^2\lambda_2\omega^2, i = 1, 2, \cdots, k.$$

It is seen that the parameter $\omega$ controls the differences between the three covariance functions. In addition, we set $a = 1.5,\ q = 11$ and $\rho = 0.1, 0.5, 0.9$ to consider the three cases when the simulated functional data have high, moderate and low correlations. We generate independent samples with three cases of the sample size vector: $\mathbf{n}_1 = [20, 30, 30]$, $\mathbf{n}_2 = [30, 40, 50]$ and $\mathbf{n}_3 = [80, 70, 100]$, representing the small, medium and large sample size cases respectively, and specify the number of design time points $J = 80$. Finally, we consider two cases of the distribution of the i.i.d. random variables $z_{ijr},\ r = 1, 2, \cdots, q;\ j = 1, 2, \cdots, n_i;\ i = 1, 2, \cdots, k$: $z_{ijr} \overset{i.i.d.}{\sim} N(0, 1)$ and $z_{ijr} \overset{i.i.d.}{\sim} t_4/\sqrt{2}$, allowing to generate Gaussian and non-Gaussian functional data respectively with $z_{ijr}$ having mean 0 and variance 1. Notice that the $t_4/\sqrt{2}$ distribution is chosen since it has nearly the heaviest tails among the $t$-distributions with finite first two moments.

For a given model configuration, the $k = 3$ groups of functional samples are generated from the data generating model (3.1). The $p$-values of $L^2_{br}, L^2_{rp}, T_{\max,rp}, \text{GPF}_{nv}, \text{GPF}_{rp}$, and $F_{\max,rp}$ are then computed. The p-value of $\text{GPF}_{nv}$ is based on the Welch-Satterthwaite $\chi^2$-approximation as given in (2.25). To compute the associated parameters $\hat{\beta}$ and $\hat{d}$, we need the estimation of $\varpi$ which is defined in (2.12). We use (2.13) instead of (2.14) in the simulations as (2.13) gives similar results to (2.14) for Gaussian data and the former can also be used for non-Gaussian data. The $p$-values of $L^2_{rp}, T_{\max,rp}$ and $F_{\max,rp}$ are obtained via using 500 runs of random permutations. The null hypothesis is rejected if the calculated $p$-value of a testing procedure is smaller than the nominal significance level $\alpha = 5\%$. We repeat the above process for 10000 times. The empirical sizes or powers of the testing procedures can then be obtained as the percentages of rejection in the 10000 runs.

### 3.2 Simulation 1

In Simulation 1, we aim to check whether the random permuted null pdfs of $\mathrm{GPF}_{rp}$ and $F_{\max,rp}$ approximate their true null pdfs well. We compare the curves of the simulated null pdfs and the first 50 random permuted null pdfs of $\mathrm{GPF}_{rp}$ and $F_{\max,rp}$ under two cases when $z_{ijr}$ , $r = 1, 2, \cdots, q;\ j = 1, 2, \cdots, n_i;\ i = 1, 2, \cdots, k$ : $z_{ijr} \overset{i.i.d.}{\sim} N(0, 1)$ and when $z_{ijr} \overset{i.i.d.}{\sim} t_4/\sqrt{2}$. For space saving, we only consider the small and large sample sizes (later we will also find that the sample sizes have little effect on the shapes of the curves). Figure 1 displays the simulated null pdfs (wider solid curves) and the 50 random permuted null pdfs (dashed curves) of $\mathrm{GPF}_{rp}$ (left 6 panels) and $F_{\max,rp}$ (right 6 panels). Note that the simulated null pdf of a testing procedure is computed using a kernel density estimator (KDE) with a Gaussian kernel based on the simulated 10000 test statistics when the null hypothesis is satisfied and a random permuted null pdf of a testing procedure is based on 10000 random permuted test statistics. The associated bandwidths are chosen automatically with the KDE software. It is seen that the random permuted null pdfs of $\mathrm{GPF}_{rp}$ and $F_{\max,rp}$ work well in approximating their underlying null pdfs under the Gaussian case.

Figure 2 displays the simulated null pdfs and the first 50 random permuted null pdfs of $\mathrm{GPF}_{rp}$ and $F_{\max,rp}$ when $z_{ijr}$ $r = 1, 2, \cdots, q;\ j = 1, 2, \cdots, n_i;\ i = 1, 2, \cdots, k$ are i.i.d. $t_4/\sqrt{2}$. It is seen that the random permutation method works generally well for $\mathrm{GPF}_{rp}$ and $F_{\max,rp}$ but not as well as when $z_{ijr}$ , $r = 1, 2, \cdots, q;\ j = 1, 2, \cdots, n_i;\ i = 1, 2, \cdots, k$ are i.i.d. $N(0, 1)$. It is seen that both Figures 1 and 2 indicate that the decay rates of the variance components $\lambda_r$, $r = 1, 2, \cdots, q$ have a great effect on the shapes of the null pdf curves of $\mathrm{GPF}_{rp}$ and $F_{\max,rp}$ while the sample sizes have little effect on them.

### 3.3 Simulation 2

In Simulation 2, we aim to compare $\mathrm{GPF}_{nv}, \mathrm{GPF}_{rp}$ and $F_{\max,rp}$ against $L^2_{br}, L^2_{rp}$ and $T_{\max,rp}$. Tables 1 and 2 present the empirical sizes and powers (in percentages) of $L^2_{br}$, $L^2_{rp}$, $T^2_{\max,rp}$, $\mathrm{GPF}_{nv}$, $\mathrm{GPF}_{rp}$ and $F_{\max,rp}$ when the $k$ functional samples follow Gaussian and non-Gaussian distributions, respectively.

First of all, it is seen that in terms of size controlling, $F_{\max,rp}$ works reasonably well under various simulation configurations while $\mathrm{GPF}_{nv}$ and $\mathrm{GPF}_{rp}$ work well only when the functional data are highly correlated or when the sample sizes are large. When the functional data are less correlated or when the sample sizes are too small, the empirical sizes of $\mathrm{GPF}_{nv}$ are too large (for Gaussian functional data) or too small (for non-Gaussian functional data) compared with the nominal size 5% and those of $\mathrm{GPF}_{rp}$ are too large for both Gaussian and non-Gaussian functional data. On the other hand, $L^2_{br}$ performs quite well under the Gaussian case but it does not work for non-Gaussian data, $L^2_{rp}$ performs well when the functional data are highly correlated or the sample sizes are large but it is liberal when the functional data are less correlated or when the sample sizes are too small, and $T_{\max,rp}$ is good under various simulation configurations. In summary, in terms of size controlling, it seems $F_{\max,rp}$ and $T_{\max,rp}$ perform similarly

Figure 1: The simulated null pdfs (wider solid curves) and the first 50 random permuted null pdfs (dashed curves) of $\mathrm{GPF}_{rp}$ and $F_{\max,rp}$ when $z_{ijr}, r = 1, \cdots, q; j = 1, \cdots, n_i; i = 1, \cdots, k \overset{i.i.d.}{\sim} N(0, 1)$.

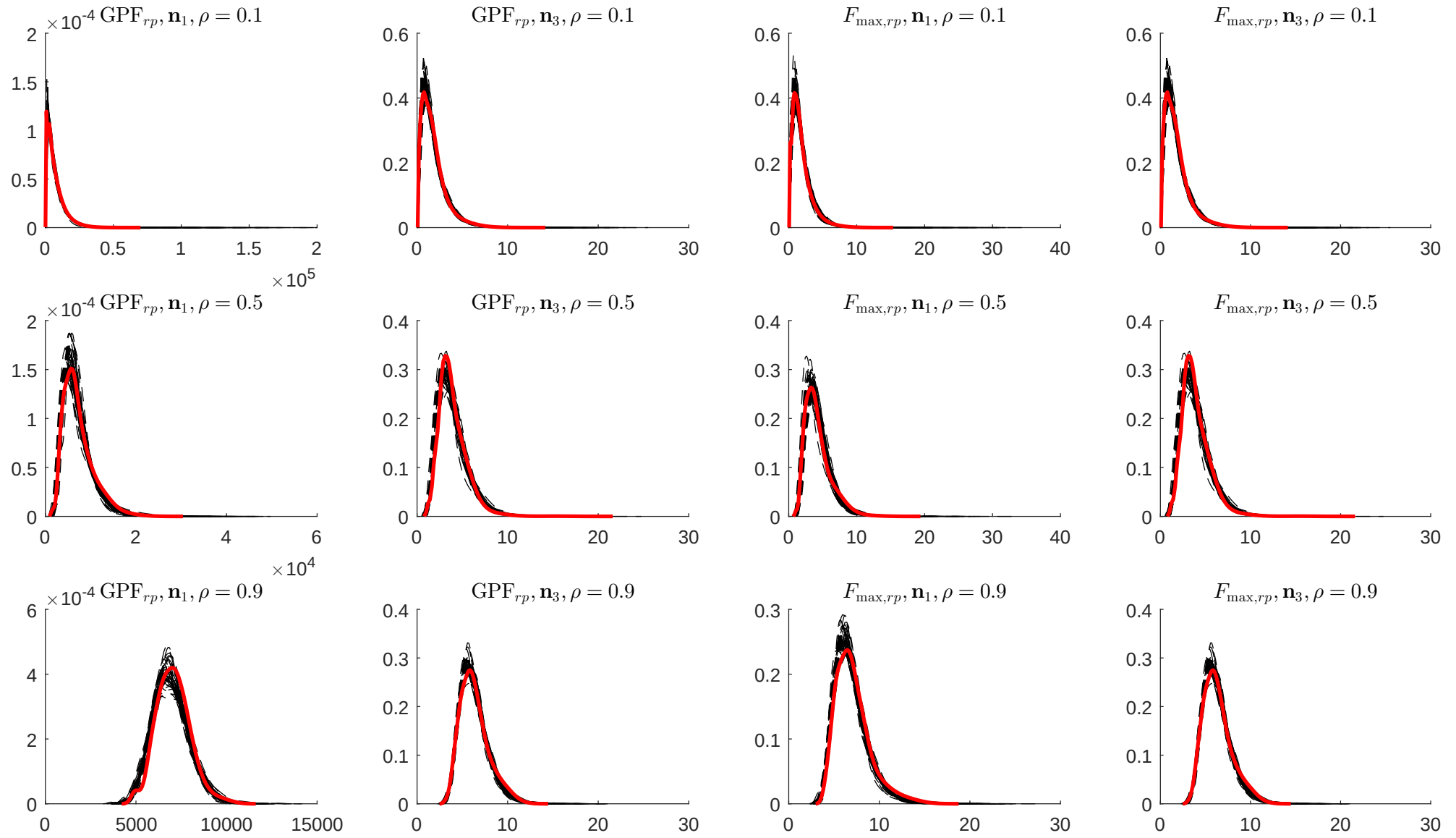

Figure 2: The simulated null pdfs (wider solid curves) and the first 50 random permuted null pdfs (dashed curves) of $\mathrm{GPF}_{rp}$ and $F_{\max,rp}$ when $z_{ijr}, r = 1, \cdots, q; j = 1, \cdots, n_i; i = 1, \cdots, k \overset{i.i.d.}{\sim} t_4/\sqrt{2}$.

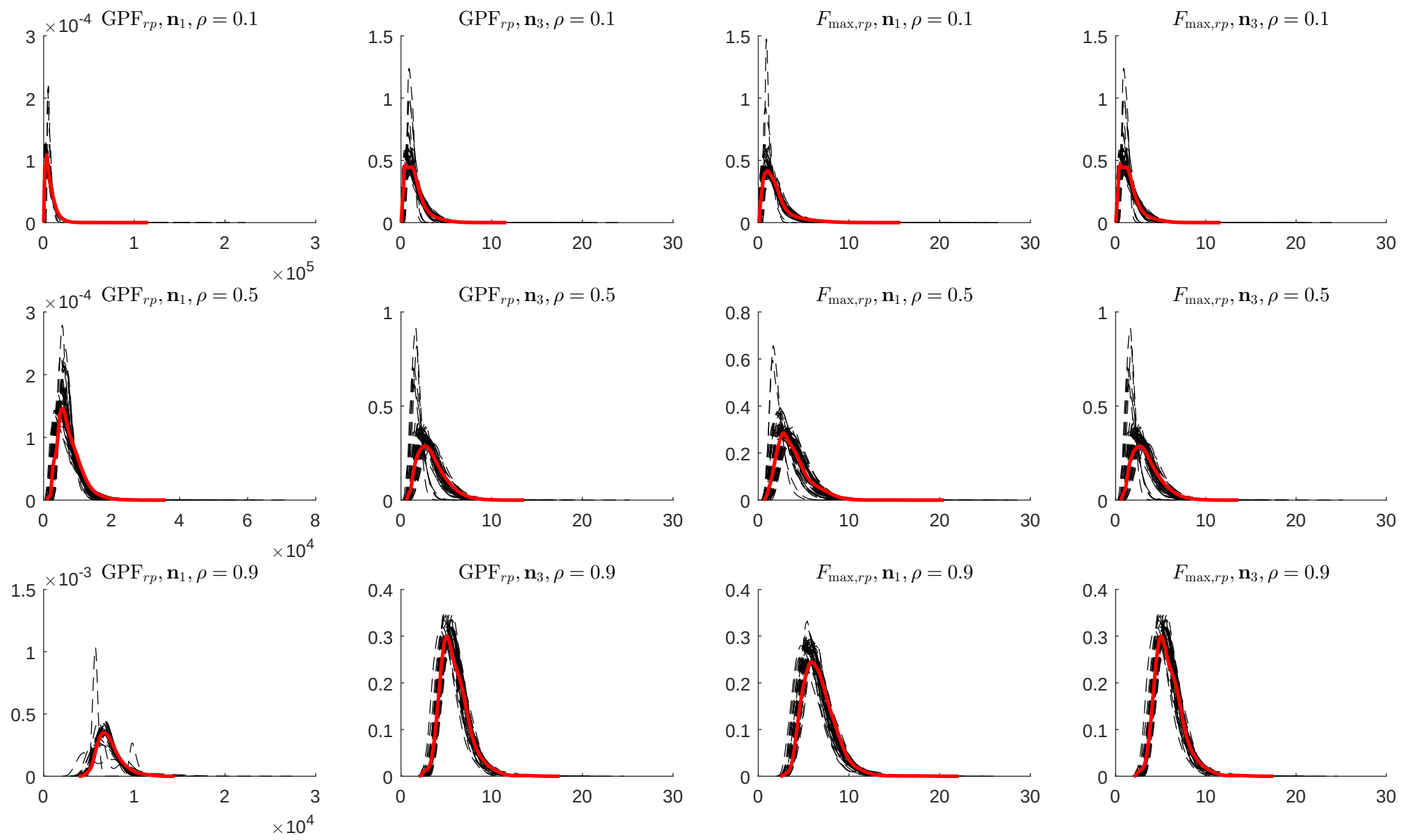

Table 1: Empirical sizes and powers (in percentages) of $L^2_{br}$, $L^2_{rp}$, $T_{\max,rp}$, $\text{GPF}_{nv}$, $\text{GPF}_{rp}$ and$F_{\max,rp}$ when $z_{ijr}, r = 1, \cdots, q; j = 1, \cdots, n_i; i = 1, \cdots, k$ are i.i.d. $N(0, 1)$.

| | $\mathbf{n}_1 = [20, 30, 30]$ | | | | | | | $\mathbf{n}_2 = [30, 40, 50]$ | | | | | | | $\mathbf{n}_3 = [80, 70, 100]$ | | | | | | |
|---|---|---|---|---|---|---|---|---|---|---|---|---|---|---|---|---|---|---|---|---|---|
| $\rho$ | $\omega_0$ | $L^2_{br}$ | $L^2_{rp}$ | $T_{\max,rp}$ | $\text{GPF}_{nv}$ | $\text{GPF}_{rp}$ | $F_{\max,rp}$ | $\omega_0$ | $L^2_{br}$ | $L^2_{rp}$ | $T_{\max,rp}$ | $\text{GPF}_{nv}$ | $\text{GPF}_{rp}$ | $F_{\max,rp}$ | $\omega_0$ | $L^2_{br}$ | $L^2_{rp}$ | $T_{\max,rp}$ | $\text{GPF}_{nv}$ | $\text{GPF}_{rp}$ | $F_{\max,rp}$ |
| | 0.0 | 4.55 | 5.74 | 5.51 | 6.34 | 5.29 | 5.16 | 0.0 | 4.63 | 5.44 | 5.38 | 5.83 | 5.08 | 5.19 | 0.0 | 4.68 | 5.13 | 4.97 | 5.15 | 5.02 | 4.88 |
| | 1.0 | 10.91 | 12.88 | 19.42 | 12.02 | 10.51 | 13.85 | 1.0 | 15.89 | 16.86 | 28.99 | 14.99 | 13.72 | 20.17 | 0.5 | 7.94 | 8.41 | 19.46 | 7.76 | 7.53 | 15.00 |
| 0.1 | 2.0 | 54.61 | 53.90 | 58.66 | 51.13 | 48.90 | 50.19 | 1.5 | 45.56 | 44.65 | 58.55 | 40.11 | 38.33 | 46.54 | 1.0 | 37.32 | 37.06 | 6545 | 31.22 | 30.69 | 53.02 |
| | 3.0 | 90.22 | 87.60 | 87.38 | 85.84 | 85.51 | 85.25 | 2.0 | 77.89 | 75.57 | 83.01 | 71.75 | 70.36 | 75.25 | 1.5 | 87.38 | 86.48 | 95.99 | 81.62 | 80.78 | 92.19 |
| | 6.0 | 99.99 | 99.67 | 99.47 | 99.33 | 99.70 | 99.59 | 3.0 | 99.01 | 98.23 | 98.56 | 97.71 | 97.70 | 98.03 | 2.0 | 99.42 | 99.19 | 99.89 | 98.64 | 98.56 | 99.67 |
| | 0.0 | 4.80 | 5.98 | 5.71 | 7.16 | 5.31 | 5.49 | 0.0 | 4.77 | 5.63 | 4.77 | 6.28 | 5.25 | 5.33 | 0.0 | 5.04 | 5.20 | 5.62 | 5.71 | 5.07 | 5.32 |
| | 0.5 | 23.24 | 25.70 | 20.03 | 25.52 | 21.00 | 18.44 | 0.4 | 24.47 | 25.91 | 20.65 | 23.82 | 20.46 | 18.74 | 0.3 | 36.64 | 36.38 | 32.82 | 31.21 | 28.90 | 26.59 |
| 0.5 | 1.0 | 74.56 | 71.72 | 56.15 | 71.54 | 68.53 | 58.71 | 0.8 | 80.66 | 78.55 | 63.48 | 76.88 | 74.09 | 63.39 | 0.4 | 62.49 | 61.91 | 53.56 | 55.47 | 53.10 | 48.22 |
| | 1.5 | 96.42 | 94.11 | 82.38 | 93.60 | 93.30 | 88.97 | 1.0 | 93.48 | 91.58 | 79.88 | 90.73 | 89.19 | 82.71 | 0.5 | 83.95 | 83.57 | 73.30 | 79.07 | 77.23 | 71.45 |
| | 2.0 | 99.65 | 98.67 | 93.33 | 98.19 | 98.41 | 97.59 | 1.2 | 98.31 | 97.46 | 90.44 | 96.97 | 96.59 | 93.58 | 0.7 | 98.65 | 98.28 | 94.44 | 98.21 | 97.81 | 95.46 |
| | 0.0 | 4.78 | 9.21 | 5.42 | 7.78 | 7.73 | 5.95 | 0.0 | 5.07 | 7.79 | 5.22 | 6.70 | 6.93 | 5.60 | 0.0 | 5.01 | 6.21 | 5.29 | 5.27 | 5.63 | 5.39 |
| | 0.5 | 32.58 | 38.42 | 11.64 | 36.62 | 34.58 | 13.11 | 0.4 | 32.68 | 36.58 | 10.52 | 32.90 | 32.40 | 12.17 | 0.2 | 19.88 | 22.17 | 8.97 | 19.02 | 20.14 | 9.02 |
| 0.9 | 1.0 | 89.47 | 87.96 | 37.29 | 89.53 | 87.89 | 44.26 | 0.5 | 52.41 | 55.27 | 14.63 | 51.69 | 50.64 | 17.75 | 0.3 | 48.49 | 50.83 | 15.28 | 45.45 | 46.19 | 16.45 |
| | 1.5 | 99.58 | 98.64 | 67.57 | 98.84 | 98.80 | 82.19 | 0.7 | 85.68 | 85.05 | 29.51 | 83.77 | 82.70 | 33.12 | 0.4 | 79.40 | 79.85 | 27.13 | 76.30 | 76.51 | 30.67 |
| | 2.0 | 100.00 | 99.75 | 84.64 | 99.70 | 99.81 | 95.96 | 1.0 | 99.23 | 98.70 | 58.28 | 98.83 | 98.53 | 68.42 | 0.5 | 95.72 | 95.57 | 43.69 | 94.37 | 94.41 | 49.82 |

Table 2: Empirical sizes and powers (in percentages) of $L^2_{br}$, $L^2_{rp}$, $T_{\max,rp}$, $\text{GPF}_{nv}$, $\text{GPF}_{rp}$ and$F_{\max,rp}$ when $z_{ijr}, r = 1, \cdots, q; j = 1, \cdots, n_i; i = 1, \cdots, k$ are i.i.d. $t_4/\sqrt{2}$.

| | $\mathbf{n}_1 = [20, 30, 30]$ | | | | | | | $\mathbf{n}_2 = [30, 40, 50]$ | | | | | | | $\mathbf{n}_3 = [80, 70, 100]$ | | | | | | |
|---|---|---|---|---|---|---|---|---|---|---|---|---|---|---|---|---|---|---|---|---|---|
| $\rho$ | $\omega_0$ | $L^2_{br}$ | $L^2_{rp}$ | $T_{\max,rp}$ | $\text{GPF}_{nv}$ | $\text{GPF}_{rp}$ | $F_{\max,rp}$ | $\omega_0$ | $L^2_{br}$ | $L^2_{rp}$ | $T_{\max,rp}$ | $\text{GPF}_{nv}$ | $\text{GPF}_{rp}$ | $F_{\max,rp}$ | $\omega_0$ | $L^2_{br}$ | $L^2_{rp}$ | $T_{\max,rp}$ | $\text{GPF}_{nv}$ | $\text{GPF}_{rp}$ | $F_{\max,rp}$ |
| | 0.0 | 30.84 | 6.40 | 5.82 | 4.78 | 5.87 | 5.45 | 0.0 | 34.14 | 5.32 | 5.09 | 4.03 | 5.10 | 5.13 | 0 | 41.13 | 5.36 | 5.25 | 3.78 | 5.20 | 5.23 |
| | 2.0 | 62.56 | 33.58 | 34.28 | 28.80 | 30.75 | 31.94 | 1.5 | 61.31 | 26.62 | 31.83 | 21.49 | 23.12 | 27.39 | 1.2 | 71.46 | 32.15 | 49.33 | 24.64 | 27.48 | 41.77 |
| 0.1 | 3.0 | 85.50 | 59.31 | 55.44 | 52.91 | 57.08 | 57.25 | 2.2 | 83.76 | 52.93 | 54.78 | 45.94 | 49.32 | 52.24 | 1.5 | 85.14 | 51.80 | 66.13 | 43.02 | 46.54 | 59.25 |
| | 4.0 | 94.90 | 73.86 | 69.14 | 66.73 | 72.58 | 72.27 | 3.0 | 96.03 | 75.94 | 73.79 | 68.66 | 73.82 | 74.79 | 2.2 | 98.01 | 85.77 | 90.21 | 79.67 | 83.69 | 88.64 |
| | 12.0 | 100.00 | 90.22 | 87.87 | 82.25 | 89.26 | 90.09 | 5.0 | 99.79 | 90.81 | 88.24 | 84.40 | 90.38 | 90.63 | 3.0 | 99.89 | 95.92 | 96.80 | 92.08 | 95.38 | 96.81 |
| | 0.0 | 42.03 | 7.33 | 6.36 | 5.17 | 6.38 | 6.07 | 0.0 | 46.43 | 6.65 | 5.62 | 4.45 | 5.76 | 5.77 | 0.0 | 54.64 | 5.40 | 5.76 | 3.57 | 5.19 | 5.34 |
| | 0.5 | 54.88 | 17.21 | 12.73 | 13.33 | 15.03 | 14.79 | 0.5 | 67.32 | 21.19 | 16.10 | 16.07 | 18.68 | 19.65 | 0.5 | 91.99 | 45.35 | 38.84 | 34.67 | 39.86 | 44.12 |
| 0.5 | 1.0 | 79.47 | 46.38 | 30.47 | 44.60 | 46.65 | 41.89 | 1.0 | 92.35 | 60.14 | 42.36 | 57.16 | 60.55 | 57.80 | 0.7 | 98.40 | 73.28 | 63.20 | 64.75 | 70.43 | 72.57 |
| | 2.0 | 97.82 | 79.06 | 61.10 | 74.07 | 80.69 | 79.25 | 1.8 | 99.69 | 88.03 | 74.16 | 83.33 | 88.89 | 88.70 | 1.0 | 99.83 | 91.67 | 85.22 | 86.28 | 90.51 | 91.65 |
| | 6.0 | 99.99 | 90.51 | 85.38 | 82.76 | 90.09 | 90.94 | 2.5 | 99.94 | 92.71 | 85.24 | 87.79 | 93.34 | 93.84 | 2.5 | 100.00 | 99.13 | 98.47 | 96.59 | 99.24 | 99.34 |
| | 0.0 | 68.04 | 10.98 | 6.12 | 3.96 | 8.67 | 6.57 | 0.0 | 73.66 | 8.54 | 5.48 | 2.97 | 7.35 | 5.91 | 0.0 | 83.98 | 6.45 | 5.33 | 2.05 | 5.99 | 5.69 |
| | 0.5 | 79.61 | 22.90 | 8.54 | 12.96 | 21.44 | 11.26 | 0.5 | 89.89 | 27.36 | 8.94 | 16.71 | 27.03 | 12.73 | 0.3 | 95.04 | 22.56 | 10.22 | 12.05 | 22.74 | 12.79 |
| 0.9 | 1.0 | 94.28 | 56.86 | 20.40 | 52.07 | 61.51 | 30.17 | 0.9 | 98.43 | 67.19 | 23.64 | 61.97 | 71.71 | 36.13 | 0.5 | 99.48 | 62.54 | 22.64 | 46.79 | 63.64 | 29.98 |
| | 1.5 | 99.01 | 77.05 | 38.33 | 75.11 | 81.46 | 59.00 | 1.2 | 99.63 | 81.67 | 38.86 | 79.59 | 86.69 | 60.48 | 0.7 | 99.93 | 87.17 | 44.90 | 79.49 | 90.45 | 58.16 |
| | 4.0 | 100.00 | 90.09 | 77.15 | 83.39 | 90.49 | 90.18 | 2.5 | 99.97 | 94.00 | 80.05 | 89.44 | 94.93 | 94.05 | 2.0 | 100.00 | 99.05 | 97.31 | 96.91 | 99.27 | 99.24 |

while $\text{GPF}_{nv}, \text{GPF}_{rp}$ and $L^2_{br}, L^2_{rp}$ perform similarly. In terms of powers, it seems $\text{GPF}_{nv}, \text{GPF}_{rp}$ and $L^2_{br}, L^2_{rp}$ have comparable powers but they have smaller (or higher) powers than $F_{\max,rp}$ and $T_{\max,rp}$ when the functional data are highly (or less) correlated.

### 3.4 Simulation 3

In Simulation 3, we aim to demonstrate that in some situations, the quasi pointwise $F$-test based tests such as $\text{GPF}_{nv}, \text{GPF}_{rp}$ and $F_{\max,rp}$ can have much better performance than $L^2_{br}, L^2_{rp}$ and $T_{\max,rp}$. For this goal, we can revise the previous data generating model slightly. That is, we specify the subject-effect functions $v_{ij}(t)$, $j = 1, 2, \cdots, n_i$, $i = 1, 2, \cdots, k$ as in the following new data generating model:

$$
\begin{aligned}
y_{ij}(t) = \eta_i(t) + v_{ij}(t),\ \eta_i(t) = \mathbf{c}_i^T[1, t, t^2, t^3]^T,\ v_{ij}(t) = \mathbf{b}_{ij}^T \mathbf{\Psi}_i(t)/(t + 1/J),\ t \in [0, 1], \\
\mathbf{b}_{ij} = [b_{ij1}, b_{ij2}, \cdots, b_{ijq}]^T,\ b_{ijr} \stackrel{d}{=} \sqrt{\lambda_r} z_{ijr},\ r = 1, 2, \cdots, q;
\end{aligned}
\tag{3.2}
$$

Table 3: Empirical sizes and powers (in percentages) of $L^2_{br}$, $L^2_{rp}$, $T_{\max,rp}$, $\text{GPF}_{nv}$, $\text{GPF}_{rp}$ and$F_{\max,rp}$ when $z_{ijr}, r = 1, \cdots, q; j = 1, \cdots, n_i; i = 1, \cdots, k$ are i.i.d. $N(0, 1)$ under the new data generating model.

| | $\mathbf{n}_1 = [20, 30, 30]$ | | | | | | | $\mathbf{n}_2 = [30, 40, 50]$ | | | | | | | $\mathbf{n}_3 = [80, 70, 100]$ | | | | | | |
|---|---|---|---|---|---|---|---|---|---|---|---|---|---|---|---|---|---|---|---|---|---|
| $\rho$ | $\omega_0$ | $L^2_{br}$ | $L^2_{rp}$ | $T_{\max,rp}$ | $\text{GPF}_{nv}$ | $\text{GPF}_{rp}$ | $F_{\max,rp}$ | $\omega_0$ | $L^2_{br}$ | $L^2_{rp}$ | $T_{\max,rp}$ | $\text{GPF}_{nv}$ | $\text{GPF}_{rp}$ | $F_{\max,rp}$ | $\omega_0$ | $L^2_{br}$ | $L^2_{rp}$ | $T_{\max,rp}$ | $\text{GPF}_{nv}$ | $\text{GPF}_{rp}$ | $F_{\max,rp}$ |
| | 0.0 | 4.54 | 5.72 | 5.34 | 6.38 | 5.34 | 5.21 | 0.0 | 4.95 | 5.51 | 5.46 | 6.13 | 5.53 | 5.59 | 0.0 | 4.89 | 4.88 | 4.80 | 5.39 | 5.14 | 5.11 |
| | 5.0 | 4.90 | 5.99 | 5.49 | 25.87 | 23.38 | 72.03 | 4.0 | 4.58 | 5.46 | 5.14 | 22.48 | 21.15 | 73.30 | 3.0 | 2.80 | 3.50 | 3.60 | 16.68 | 16.38 | 69.13 |
| 0.1 | 7.0 | 4.33 | 5.35 | 5.30 | 59.39 | 56.75 | 94.90 | 5.5 | 4.81 | 5.66 | 5.37 | 50.72 | 48.76 | 97.22 | 4.5 | 4.73 | 4.93 | 4.90 | 55.14 | 53.93 | 99.64 |
| | 10.0 | 4.60 | 6.39 | 5.00 | 92.21 | 92.51 | 99.00 | 7.0 | 4.77 | 5.32 | 5.17 | 81.18 | 80.09 | 99.84 | 5.0 | 5.59 | 5.79 | 5.39 | 72.43 | 71.73 | 100.00 |
| | 14.0 | 4.70 | 5.89 | 5.00 | 99.40 | 99.80 | 99.80 | 8.5 | 5.03 | 5.79 | 5.30 | 94.93 | 94.85 | 99.97 | 7.0 | 4.00 | 4.40 | 4.50 | 98.70 | 98.50 | 100.00 |
| | 0.0 | 4.58 | 5.58 | 5.39 | 6.59 | 5.50 | 5.18 | 0.0 | 5.15 | 5.96 | 5.82 | 6.86 | 5.59 | 5.38 | 0.0 | 4.76 | 4.99 | 5.08 | 5.53 | 5.02 | 5.34 |
| | 1.0 | 4.90 | 5.39 | 4.70 | 15.98 | 12.19 | 13.19 | 1.2 | 4.49 | 5.01 | 4.81 | 23.76 | 20.33 | 24.48 | 0.9 | 4.86 | 5.30 | 5.24 | 25.21 | 23.32 | 30.94 |
| 0.5 | 2.5 | 4.20 | 5.89 | 5.39 | 58.84 | 53.85 | 61.04 | 2.0 | 4.65 | 5.66 | 5.36 | 59.41 | 55.10 | 64.86 | 1.2 | 4.80 | 4.50 | 4.90 | 44.36 | 42.36 | 54.05 |
| | 3.5 | 4.63 | 5.80 | 5.36 | 85.25 | 83.13 | 88.31 | 2.7 | 4.80 | 5.79 | 5.53 | 84.59 | 82.12 | 90.75 | 1.8 | 4.80 | 5.19 | 4.80 | 83.42 | 81.32 | 90.51 |
| | 5.0 | 5.39 | 6.89 | 6.39 | 98.20 | 98.30 | 98.60 | 3.3 | 5.27 | 5.91 | 5.60 | 94.93 | 94.06 | 97.86 | 2.2 | 4.30 | 4.50 | 5.09 | 96.10 | 95.80 | 98.50 |
| | 0.0 | 4.70 | 4.80 | 5.29 | 6.29 | 6.69 | 5.79 | 0.0 | 4.63 | 5.51 | 5.27 | 6.46 | 6.87 | 5.45 | 0.0 | 4.70 | 4.80 | 5.59 | 4.60 | 5.29 | 4.90 |
| | 1.5 | 4.40 | 5.00 | 4.30 | 34.57 | 34.17 | 11.49 | 1.0 | 4.93 | 5.74 | 5.49 | 24.72 | 25.30 | 11.22 | 0.8 | 5.89 | 5.89 | 5.29 | 30.07 | 31.47 | 12.39 |
| 0.9 | 2.0 | 4.53 | 5.69 | 5.43 | 58.87 | 57.93 | 19.76 | 1.6 | 4.65 | 5.36 | 5.05 | 56.32 | 56.81 | 20.65 | 1.0 | 5.17 | 5.63 | 5.34 | 44.08 | 45.34 | 17.51 |
| | 2.8 | 4.92 | 6.00 | 5.45 | 87.72 | 87.27 | 43.10 | 2.3 | 4.95 | 5.57 | 5.34 | 88.91 | 88.93 | 48.27 | 1.5 | 4.70 | 6.09 | 5.49 | 85.11 | 85.51 | 43.16 |
| | 3.5 | 4.80 | 5.69 | 6.39 | 96.50 | 96.40 | 70.73 | 2.8 | 4.71 | 5.72 | 5.60 | 97.41 | 97.33 | 71.89 | 2.0 | 5.89 | 5.39 | 5.89 | 99.70 | 99.70 | 72.53 |

$j = 1, 2, \cdots, n_i$, $i = 1, 2, \cdots, k$. In addition, we modify the second basis function via setting $\psi_{12}(t) = \psi_{32}(t) = \sqrt{2}\sin(2\pi t)$ and $\psi_{22}(t) = \sqrt{2}sin(2\pi t) + t\omega$. The term $t\omega$ is used to control the difference between the three covariance functions. In this new data generating model, the covariance functions have different scales at different time points. As $\text{GPF}_{nv}, \text{GPF}_{rp}$ and $F_{\max,rp}$ are scale-invariant, we expect that they should have better performance than $L^2_{br}, L^2_{rp}$ and $T_{\max,rp}$ which are not scale-invariant. This is indeed the case as shown by the simulation results presented in Table 3 where it is seen that $\text{GPF}_{nv}, \text{GPF}_{rp}$ and $F_{\max,rp}$ are more powerful than $L^2_{br}, L^2_{rp}$ and $T_{\max,rp}$ whose empirical powers are always around the nominal sizes.

# 4 A Real Data Example

In this section, we present a real data example for applications of the quasi GPF tests ($\text{GPF}_{nv}, \text{GPF}_{rp}$) and the quasi $F_{\max}$ test ($F_{\max,rp}$), together with $L^2_{br}, L^2_{rp}$ and $T_{\max,rp}$ tests. The real functional data set was collected by Professor Carey at UCD in a medfly rearing facility in Mexico. It recorded the number of alive medflies over a period of time aiming to quantify the effects of nutrition and gender on mortality. The data set was kindly made available online by Professor Hans-Georg Müller and Professor Carey's laboratory at http://anson.ucdavis.edu/~mueller/data/data.html and has been extensively studied in Müller et al. (1997) and Müller and Wang (1998).

Table 4: P-values (in percentages) of $L^2_{br}$, $L^2_{rp}$, $T_{\max,rp}$, $\text{GPF}_{nv}$, $\text{GPF}_{rp}$ and $F_{\max}$ applied to the survival functions of the four groups of medflies.

| Group Comparison | $L^2_{br}$ | $L^2_{rp}$ | $T_{\max,rp}$ | $\text{GPF}_{nv}$ | $\text{GPF}_{rp}$ | $F_{\max,rp}$ |
|---|---|---|---|---|---|---|
| Group 1 vs Group 2 | 44.52 | 24.61 | 26.26 | 19.49 | 21.61 | 6.24 |
| Group 3 vs Group 4 | 2.79 | 0.94 | 0.50 | 0.75 | 0.77 | 0.08 |
| Group 1 vs Group 3 | 15.42 | 6.15 | 0.89 | 2.57 | 3.23 | 0.14 |
| Group 2 vs Group 4 | 49.10 | 34.79 | 48.95 | 10.23 | 11.33 | 0.61 |
| All the four groups | 12.04 | 1.52 | 1.04 | 0.14 | 0.23 | 0.02 |

The data set consists of the lifetimes of four groups of medflies over 101 days. Each group has 33 cohorts with each cohort consisting of about 3000-4000 medflies. The four groups of medflies are "1. males on sugar diet", "2. males on protein plus sugar diet", "3. females on sugar diet" and "4. females on protein plus sugar diet". In applications, the cohort survival behavior can be conveniently summarized in the form of a survival function. This survival function can be obtained by dividing the daily number of alive medflies by the total number of medflies in each cohort at the beginning. For simplicity, we only consider the survival functions on the first 2-31 days since on the first day all the survival functions equal 1. It is of interest to check if the covariance structures of the four different groups of medflies are the same.

Table 4 shows the p-values (in percentages) of $L^2_{br}$, $L^2_{rp}$, $T_{\max,rp}$, $\text{GPF}_{nv}$, $\text{GPF}_{rp}$ and $F_{\max,rp}$ applied to several selected group comparisons of the survival functions of the four groups of medflies. For different group comparisons, the goals are different. The comparison "Group 1 vs Group 2" aims to assess the effect of the sugar diet on male medflies, the comparison "Group 3 vs Group 4" aims to assess the effect of the sugar diet on female medflies, the comparison "Group 1 vs Group 3" aims to assess the gender effect of the sugar diet, the comparison "Group 2 vs Group 4" aims to assess the gender effect of the protein plus sugar diet, and "All the four groups" comparison aims to test if all the four groups have the same covariance structure.

It is seen that all the p-values of the tests for the comparison of "Group 1 vs Group 2" suggest that the effect of the sugar diet on male medflies is not significant, showing that the sugar diet may be useless for male medflies. However, it is not the case for the effect of the sugar diet on female medflies since all the p-values of the tests for the comparison of "Group 3 vs Group 4" suggest that the effect of the sugar diet on female medflies is highly significant. Therefore, it is expected that the gender effect of the sugar diet should be significant and it is also expected that the gender effect of the protein plus sugar diet should be significant. However, only the p-values of $T_{\max,rp}$, $\text{GPF}_{nv}$, $\text{GPF}_{rp}$, $F_{\max,rp}$ for the comparison of "Group 1 vs Group 3" suggest that the gender effect of the sugar diet is highly significant and only the p-value of $F_{\max,rp}$ for the comparison of "Group 2 vs Group 4" suggest that the gender effect of the protein plus sugar diet is highly significant. All the P-values of the tests except $L^2_{br}$ for the comparison "All the four groups" suggest that the covariance structures of the four groups are unlikely the same. The p-values in this table suggests that the suprenum based tests such as $T_{\max,rp}$ and $F_{\max,rp}$ are more powerful than other tests, and the pointwise quasi $F$-test based tests such as $\mathrm{GPF}_{nv}, \mathrm{GPF}_{rp}$ and $F_{\max,rp}$ are generally more powerful than those $L^2$-norm based tests such as $L^2_{br}, L^2_{rp}$. It is also seen that the $F_{\max,rp}$ test is the most powerful test among all the tests under consideration.

# Appendix

Technical proofs and additional contents are available in supplementary materials.